\documentclass[a4paper]{spie}  
\usepackage[]{graphicx}
\usepackage{authblk}
\title{Simons Observatory Small Aperture Telescope overview} 
\author[a]{Kenji Kiuchi}
\author[b]{Shunsuke Adachi}
\author[c]{Aamir M. Ali}
\author[d]{Kam Arnold}
\author[c,e,f]{Peter Ashton}
\author[g]{Jason E. Austermann}
\author[h]{Andrew Bazako}
\author[g]{James A. Beall}
\author[f,i]{Yuji Chinone}
\author[j]{Gabriele Coppi}
\author[h]{Kevin D. Crowley}
\author[c]{Kevin T. Crowley}
\author[k]{Simon Dicker}
\author[g]{Bradley Dober}
\author[g]{Shannon M. Duff}
\author[l]{Giulio Fabbian}
\author[d]{Nicholas Galitzki}
\author[m]{Joseph E. Golec}
\author[n]{Jon E. Gudmundsson}
\author[m]{Kathleen Harrington}
\author[o]{Masaya Hasegawa}
\author[p]{Makoto Hattori}
\author[c,e]{Charles A. Hill}
\author[q]{Shuay-Pwu Patty Ho}
\author[g]{Johannes Hubmayr}
\author[r]{Bradley R. Johnson}
\author[o]{Daisuke Kaneko}
\author[f]{Nobuhiko Katayama}
\author[d]{Brian Keating}
\author[a,e,i,s]{Akito Kusaka}
\author[t]{Jack Lashner}
\author[c]{Adrian T. Lee}
\author[f]{Frederick Matsuda}
\author[f]{Heather McCarrick}
\author[a]{Masaaki Murata}
\author[j]{Federico Nati}
\author[a]{Yume Nishinomiya}
\author[h]{Lyman Page}
\author[e,u]{Mayuri Sathyanarayana Rao}
\author[v]{Christian L. Reichardt}
\author[a]{Kana Sakaguri}
\author[f]{Yuki Sakurai}
\author[d]{Joseph Seibert}
\author[d]{Jacob Spisak}
\author[b]{Osamu Tajima}
\author[d]{Grant P. Teply}
\author[a]{Tomoki Terasaki}
\author[d]{Tran Tsan}
\author[g]{Samantha Walker}
\author[w]{Edward J. Wollack}
\author[k,x]{Zhilei Xu}
\author[a]{Kyohei Yamada}
\author[j]{Mario Zannoni}
\author[k]{Ningfeng Zhu}

\affil[a]{Department of Physics, The University of Tokyo, 7-3-1 Hongo, Bunkyo, Tokyo 113-0033, Japan}
\affil[b]{Department of Physics, Faculty of Science, Kyoto University, Kitashirakawa Oiwake-cho, Sakyo-ku, Kyoto 606-8502, Japan}
\affil[c]{Department of Physics, University of California, Berkeley, LeConte Hall, Berkeley, CA 94720, USA}
\affil[d]{Department of Physics, University of California, San Diego, La Jolla, CA  92093, USA}
\affil[e]{Physics Division, Lawrence Berkeley National Laboratory, Berkeley, CA 94720, USA}
\affil[f]{Kavli Institute for The Physics and Mathematics of The Universe (WPI), The University of Tokyo, Kashiwa, 277- 8583, Japan}
\affil[g]{NIST Quantum Sensors Group, 325 Broadway Mailcode 687.08, Boulder, CO 80305, USA}
\affil[h]{Joseph Henry Laboratories of Physics, Jadwin Hall, Princeton University, Princeton, NJ 08544, USA}
\affil[i]{Research Center for the Early Universe, School of Science, The University of Tokyo, 7-3-1 Hongo, Bunkyo, Tokyo 113-0033, Japan}
\affil[j]{Department of Physics, University of Milano-Bicocca Piazza della Scienza, Milano (MI), Italy}
\affil[k]{Department of Physics and Astronomy, University of Pennsylvania, 209 South 33rd Street, Philadelphia, PA 19104, USA}
\affil[l]{Department of Physics and Astronomy, University of Sussex, Brighton BN1 9QH, UK}
\affil[m]{Department of Astronomy and Astrophysics, University of Chicago, 5640 South Ellis Avenue, Chicago, IL 60637, USA}
\affil[n]{Department of Physics, Stockholm University, SE-106 91 Stockholm, Sweden}
\affil[o]{Institute of Particle and Nuclear Studies, High Energy Accelerator Research Organization, 1-1 Oho, Tsukuba, Ibaraki 305-0801, Japan}
\affil[p]{Astronomical Institute, Tohoku University, 6-3 Aramaki Aza Aoba, Aoba, Sendai, Miyagi 980-8578, Japan}
\affil[q]{Department of Physics, Stanford University, 382 Via Pueblo, Stanford, CA 94305, USA}
\affil[r]{Department of Astronomy, University of Virginia, Charlottesville, VA 22903, USA}
\affil[s]{Kavli Institute for the Physics and Mathematics of the Universe (WPI), Berkeley Satellite, the University of California, Berkeley 94720, USA}
\affil[t]{Department of Physics and Astronomy, University of Southern California, Los Angeles, CA 90089, USA}
\affil[u]{Raman Research Institute, C. V. Raman Avenue, 5th Cross Road, Sadashivanagar, Near Mekhri Circle, Bengaluru, Karnataka 560080, India}
\affil[v]{School of Physics, University of Melbourne, Parkville, VIC 3010, Australia}
\affil[w]{NASA/Goddard Space Flight Center, Greenbelt, MD, USA}
\affil[x]{Kavli Institute, Massachusetts Institute of Technology, Cambridge, MA, USA}

\authorinfo{Kenji K: E-mail: kkiuchi@phys.s.u-tokyo.ac.jp}
 
\begin{document} 
\maketitle 
\begin{abstract}
The Simons Observatory (SO) is a cosmic microwave background (CMB) experiment from the Atacama Desert in Chile comprising three small-aperture telescopes (SATs) and one large-aperture telescope (LAT). In total, SO will field over 60,000 transition-edge sensor (TES) bolometers in six spectral bands centered between 27 and 280 GHz in order to achieve the sensitivity necessary to measure or constrain numerous cosmological quantities.
In this work, we focus on the SATs which are optimized to search for primordial gravitational waves that are detected as parity-odd polarization patterns called a B-modes on degree scales in the CMB. Each SAT employs a single optics tube with TES arrays operating at 100 mK. The high throughput optics system has a 42 cm aperture and a 35-degree field of view coupled to a 36 cm diameter focal plane. The optics consist of three metamaterial anti-reflection coated silicon lenses. Cryogenic ring baffles with engineered blackbody absorbers are installed in the optics tube to minimize the stray light. The entire optics tube is cooled to 1~K. A cryogenic continuously rotating half-wave plate near the sky side of the aperture stop helps to minimize the effect of atmospheric fluctuations. The telescope warm baffling consists of a forebaffle, an elevation stage mounted co-moving shield, and a fixed ground shield that together control the far side-lobes and mitigates ground-synchronous systematics.
We present the status of the SAT development.

\end{abstract}



\section{INTRODUCTION}
\label{sec:intro}  
The Simons Observatory (SO) is designed to measure the polarization from the Atacama Desert in Chile at an altitude of 5,200~m.
In order to measure large angular scale pattern to small angular scale pattern, the SO will have three small aperture telescopes (SATs) and one large aperture telescope (LAT).
The LAT has a six meter aperture and more than 30,000 transition edge sensors (TESes).
The SATs have a 42~cm aperture with more than 30,000 transition edge sensors distributed across the three SATs.\\
The SATs will measure 10\% of the full sky \cite{soforecast2019}.
Three SATs have 2 frequency bands each, for a total of 6 bands between bands with center frequency of 27 GHz to 280 GHz in order to subtract synchrotron radiation and dust emission.
The Middle Frequency (MF) has centre frequencies of 93~GHz and 145~GHz, the Ultra High Frequency (UHF) has bands centered at 225~GHz and 280~GHz, while the Low Frequency (LF) will be developed for frequency bands around 27~GHz and 39~GHz.
Three SATs will observe with two MF configuration and one UHF configuration. And if needed, one MF telescope will be replaced by LF configuration for a single year of LF observation.
\\
The SATs are optimized for measuring degree-scale B-modes produced by the primordial gravitational wave\cite{bmode1,bmode2}. Hence their target multipole range is 30 $< \ell <$ 300. The science goal is to measure the tensor to scalar ratio, r, with $\sigma(r)=0.002$ or better\cite{soforecast2019, sodecadal2019}. 
We have begun integration in 2020 and aim to deploy the first SAT in Chile in 2021.
\begin{figure}
    \centering
    \begin{tabular}{cc}
    \includegraphics[height=7cm]{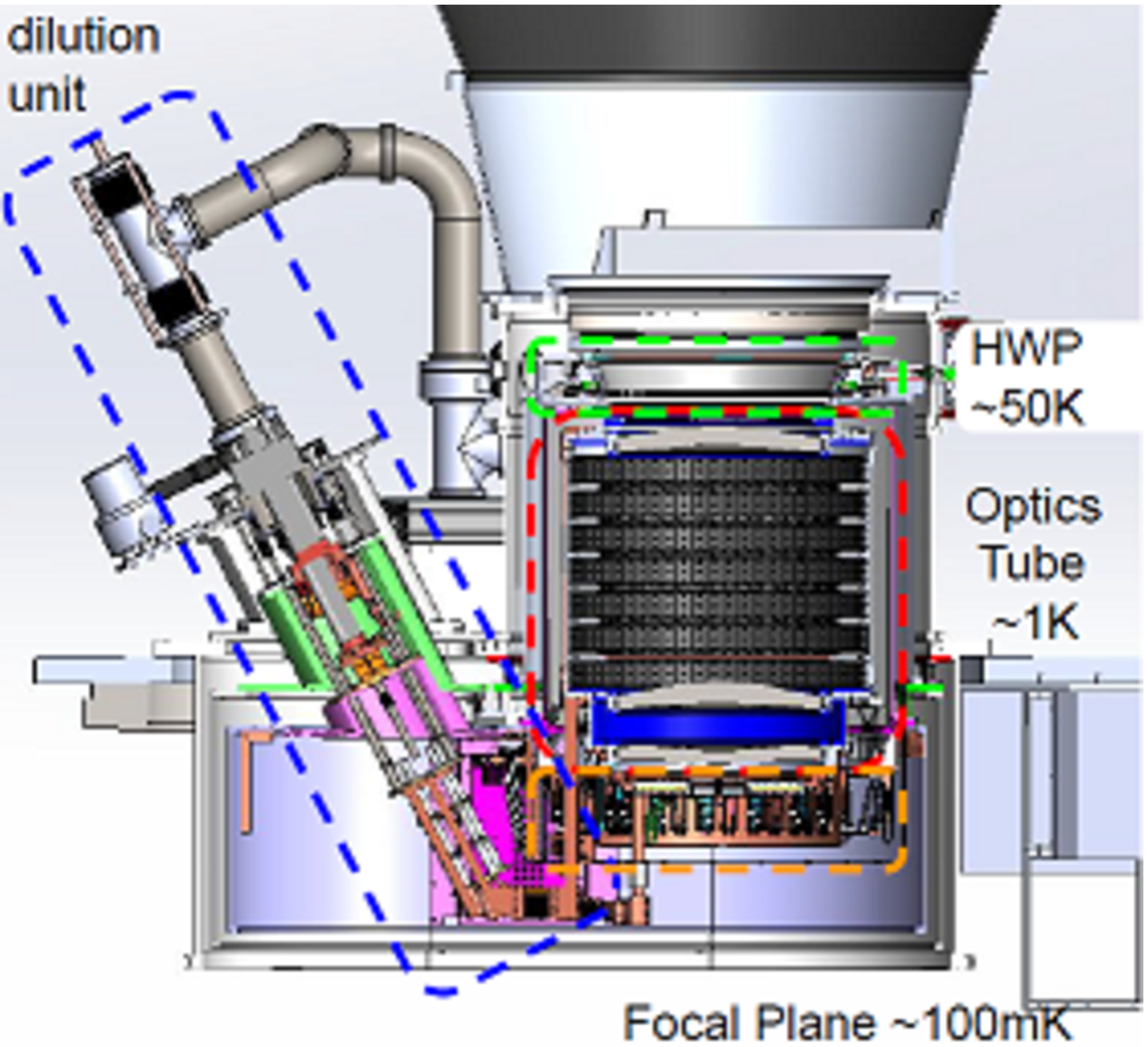}
    \includegraphics[height=7cm]{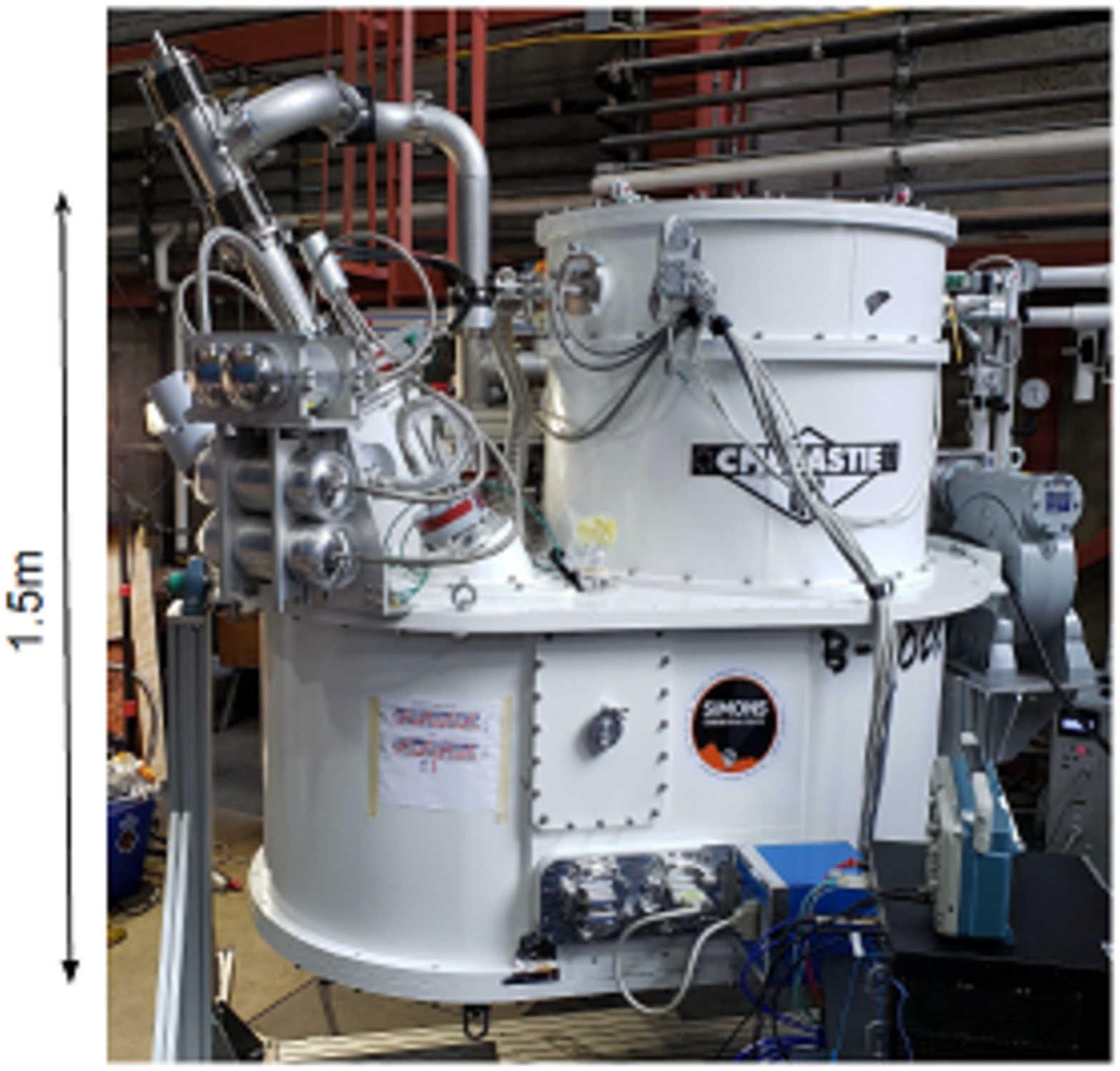}
    \end{tabular}
    \caption{The $left$ figure shows a cross section of a SAT receiver CAD model. Top side is a sky side. $Green\ dashed\ line$ shows the HWP that cooled at 50~K using the PTC. The dilution refrigerator cools the optics tube ($red\ dash\ line$) and focal plane ($orange\ dash\ line$) at 1~K and 100~mK, respectively. The dilution refrigerator and pulse tube cryocooler are tilted to be vertical when observation at an elevation of 50${}^\circ$. 
    The $right$ picture is the cryostat for the first SAT.}
    \label{fig:SAT_cryo}
\end{figure}

\section{Instruments} 
\label{sec:sections}
Figure \ref{fig:SAT_cryo} $(left)$ shows a cross section of SAT. 
The SAT cryostat consists of two refrigerators, one optics tube including three silicon lenses, focal plane with TES arrays and multiplexing chips and the half wave plate (HWP) signal modulator.
On top of the cryostat, there is a the forebaffle and a ftsparse wire-grid calibrator.
This section explains the design of the each components and its status.
\subsection{Optical design} 
The SAT optics have a 42~cm aperture and a diffraction-limited field of view (FOV) of 35${}^\circ$ couple to a 36~cm diameter focal plane
that accommodate seven detector modules. 
The primary constraint for the optics is given by the available size of optical components.
We will use the HWP polarization signal modulator for systematic mitigation.
The largest available size of sapphire that has the property we desire is 51~cm in diameter.
We employ lenses made by single crystalline silicon with meta-material AR cut on surfaces\cite{datta2013ApOpt..52.8747D}.
The largest available size of a silicon ingot for the lenses is 46~cm in diameter.
The optics is optimized with a three-lens configuration with these constraints.
The SAT optics realize high strehl ratio of $>0.89$.
\subsubsection{Warm baffle component}
The SAT has a three component baffling scheme to suppress the ground pick up. 
A forebaffle is mounted on the cryostat and is able to move with entire optics.
The second component is a comoving shield mounted on azimuth stage.
The third component is a ground screen. The radius and height of ground shield is 8.2~m and 5.6~m, respectively.
The combination of three baffles blocks the direct ray path to nearby mountains and mitigates ground synchronous pick up.
\begin{figure}
    \centering
    \begin{tabular}{cc}
    \includegraphics[height=6.5cm]{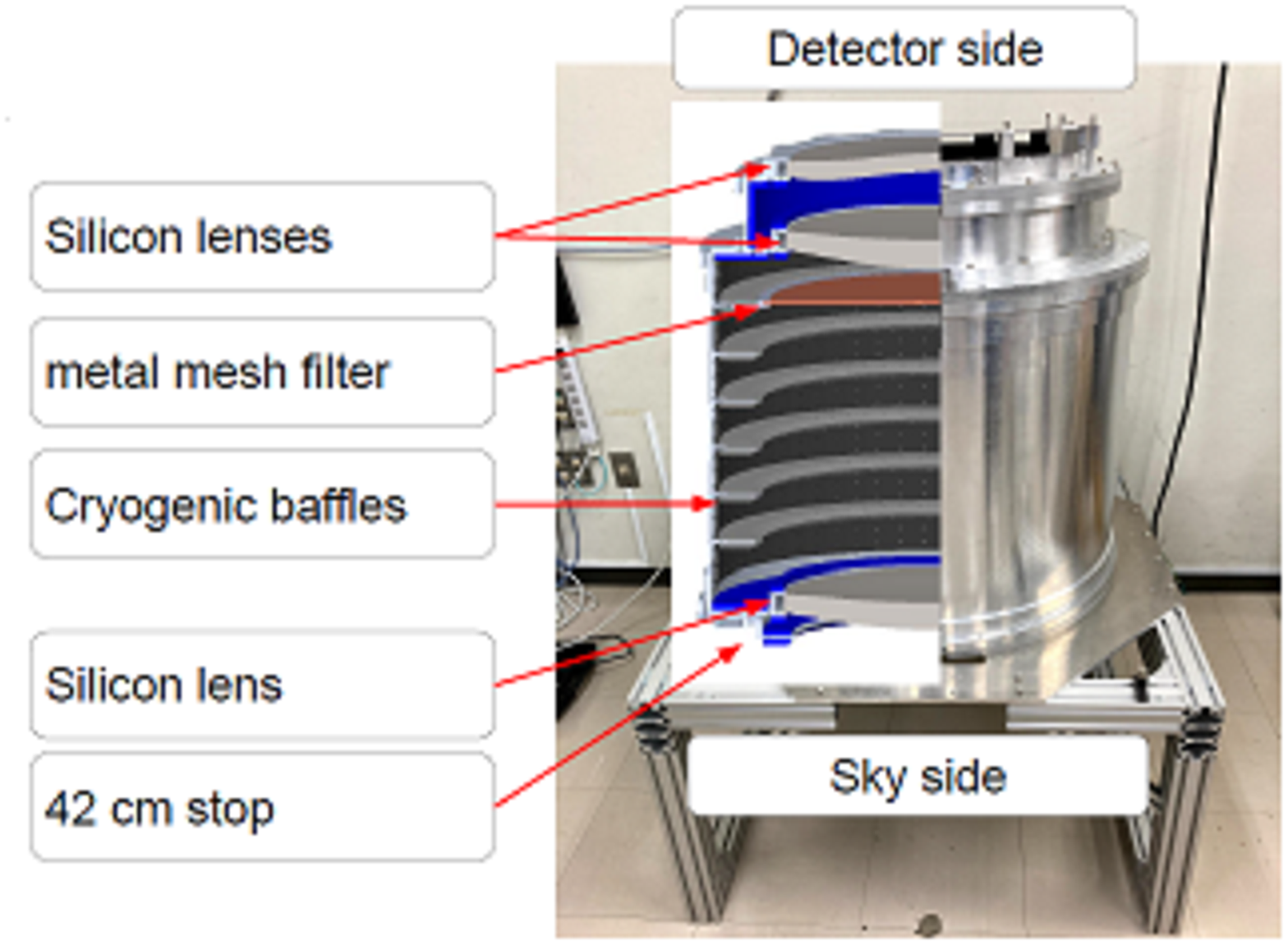}
    \includegraphics[height=6.5cm]{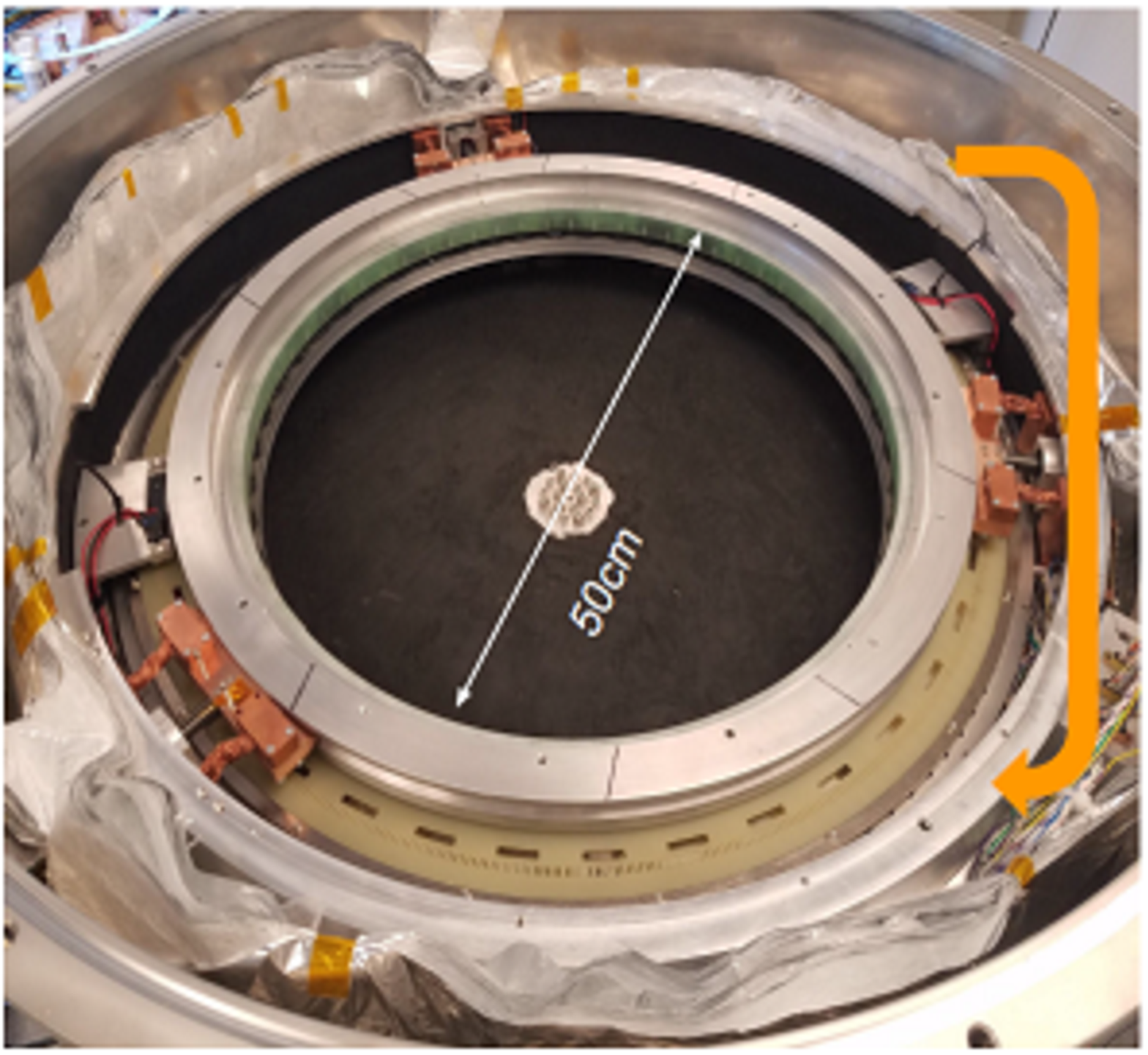}
    \end{tabular}
    \caption{$Left$ picture shows a cross section of 3D CAD model of optics tube (OT) and the picture of fabricated optics tube. $Right$ picture is a half-wave plate rotation mechanism.}
    \label{fig:ot_hwp}
\end{figure}

\subsection{Cryostat and Platform} 
Figure \ref{fig:SAT_cryo} $(right)$ shows a picture of a SAT cryostat.
Each cryostat has a dilution refrigerator (DR, BlueFors BF-SD400) including one pulse tube cryocooler (PTC) and one additional PTC (Cryomech PT420).
The focal plane is cooled to 100~mK via the mixing chamber stage of dilution refrigerator.
The optics tube including lenses is cooled to 1~K by the still stage of dilution refrigerator.
The additional PTC cools the 50K radiation shell and 4K radiation shell.
The HWP is mounted on the 50K shell.
The DR provides 400 $\mu$W cooling power at 100~mK and allow us to operate $>$10,000 sensors in one SAT.
Thermal loading to each stage is well controlled{\cite{nick2018}}.
The SAT platform has azimuth, elevation, boresight rotational degree of freedom.
The boresight rotation is limited to $\pm$ 90${^\circ}$ due to the limitation of the cooling power of PTC.
\subsection{Optics Tube} 
Figure \ref{fig:ot_hwp} $(left)$ shows a picture and a cross section of optics tube.
The optics tube (OT) for first SAT is built and has been integrated into SAT cryostat.
The SAT optics tube consists three silicon lenses, lens mounts, cryogenic baffles, metal mesh filters and the 42~cm cold stop.
\subsubsection{Mechanical and Thermal}
The entire structure operates at 1~K. This low temperature suppresses radiative loading on the detector.
The mechanical structure including lens mounts are made of pure aluminum ($>$ 99.5\%) to ensure good thermal conductivity.
Five machined parts support lenses with a position accuracy of $<$ 0.5~mm.
The coefficient of the thermal expansion of pure aluminum is well known and leads to a 0.4\% contraction from room temperature to 1~K.
The AR coating is required to be transparent (low reflectance and low loss) across a wide-band and have a matched coefficient of thermal expansion (CTE).
Our AR coating is made of silicon itself using silicon dicing technique \cite{datta2013ApOpt..52.8747D,Joey2020}.
This technology is already used ACTPol and satisfies all our requirements.
Segments of compressed metal RF gaskets\footnote{spira gasket: https://www.spira-emi.com/} are used for the lens mount to absorb CTE mismatch between the Si lens and Al mount.
The optics tube will be cooled at 1~K level via a thermal strap connected to the still stage of the dilution unit.
The thermal conductivity of the aluminum is degraded below the superconducting transition temperature of 1.2~K,
while the expected temperature of the OT is slightly higher than that.
Pure aluminum has a good balance of thermal conductivity, weight and machinability.
The inner edge of the cold stop has double side taper with taper angle of 40 degrees.
\subsubsection{Cryogenic Baffling}
The cryogenic baffles consist of six ring shape discs, with the last ring being used as a mount for metal-mesh low-pass filters.
The cryogenic baffles are designed to block bounces (reflection and scattering) from sky-side to detector side.
The inner edge of cryogenic baffles have 10~$\lambda$ clearance to the beam at 150~GHz and its shape is double side knife edge to suppress reflection at the edges.
The inner surface of the optics tube is blackened using two types of blackbody absorbers.
One is fabricated using injection molding technique\cite{xu2020arXiv201002233X}.
The carbon loaded plastic is formed to black tiles that have pyramid structure.
Injection molding is suitable for mass production and $>$ 500 pieces of blackbodies covers cryogenic baffling section where is a large surface area in one OT.
This blackbody has CTE mismatch compared to the aluminum structure, we screwed on aluminum plate with suitable torque to avoid damage due to thermal contraction.
A second types of blackbody is fabricated using 3D printing technique\cite{adachi2020}.
The 3D pieces are suitable for covering relatively complex surface like lens mounts.
We fabricated plastic pyramidal mold using a 3D printer and the mold is filled with stainless-steel loaded Stycast 2850FT.
This blackbody is glued using Stycast 2850FT because of the similar CTE to the aluminum.
The reflectances of both blackbodies are expected to be $<$ 1\%. Mounting methods are also expected to have good thermal conductivity at the 1~K.
Combined with new blackbodies and baffling structure, the OT minimizes stray light to the detector.
This will improve the sensitivity and mitigate the systematics.
\subsection{Cryogenic half wave plate (CHWP)} 
Figure \ref{fig:ot_hwp} $(right)$ picture shows the HWP rotation mechanism for the first SAT.
\subsubsection{Optical}
The cryogenic continuously rotating half wave plate (HWP) rapidly modulates the polarization signal at a higher frequency.
The target rotational speed of our HWP is 2~Hz and this polarization modulation frequency will be 8~Hz above the $1/f$ fluctuations of the atmosphere.
Another advantage of the HWP is a result of the rotation of the transmitted polarization.
A TES, which is sensitive to polarization in a single direction, is able to measure polarization in two orthogonal directions due to the rotating effect of the spinning HWP.
\subsubsection{Mechanical}
The mechanical parts are already fabricated and tested in a SAT cryostat.
The design of the HWP system is based on the POLARBEAR-2b HWP system\cite{charlie2020arXiv200903972H}.
The HWP system is mounted on the 50~K stage of pulse tube cryocooler (PTC).
The PTC has sufficient cooling power at this stage and also the radiation from the sapphire is already suppressed at this temperature.
The optical diameter of HWP will limit the aperture size of optics.
We employed three 50.5~cm diameter sapphire disc stacked with achromatic combination.
The clear aperture of the HWP system is 478~mm.
The superconducting magnetic bearing shows low power dissipation at 50K and the HWP can rotate continuously during observation period.
\subsection{Sparse wire-grid calibrator} 
\begin{figure}
    \centering
    \begin{tabular}{cc}
    \includegraphics[height=7cm]{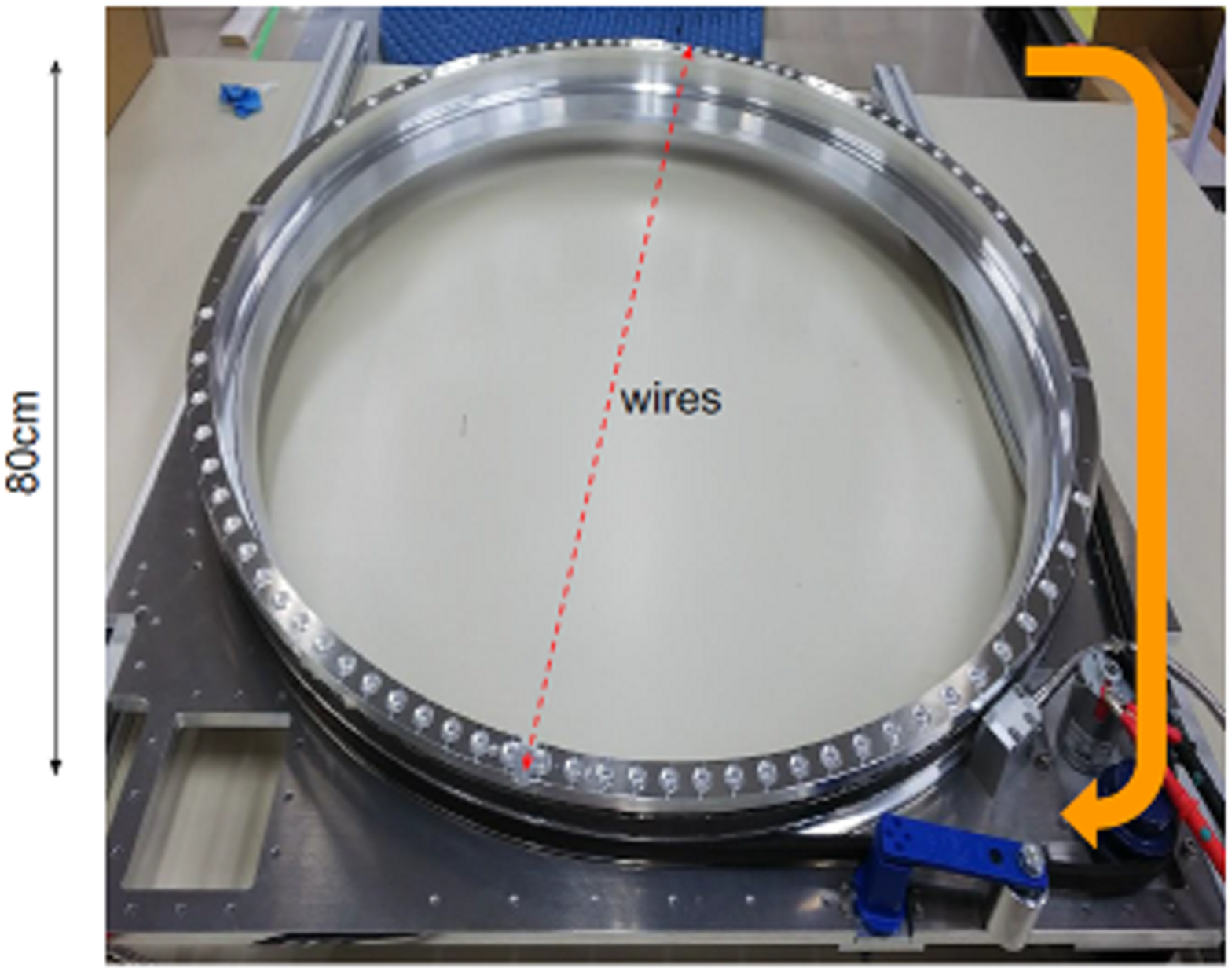}
    \includegraphics[height=7cm]{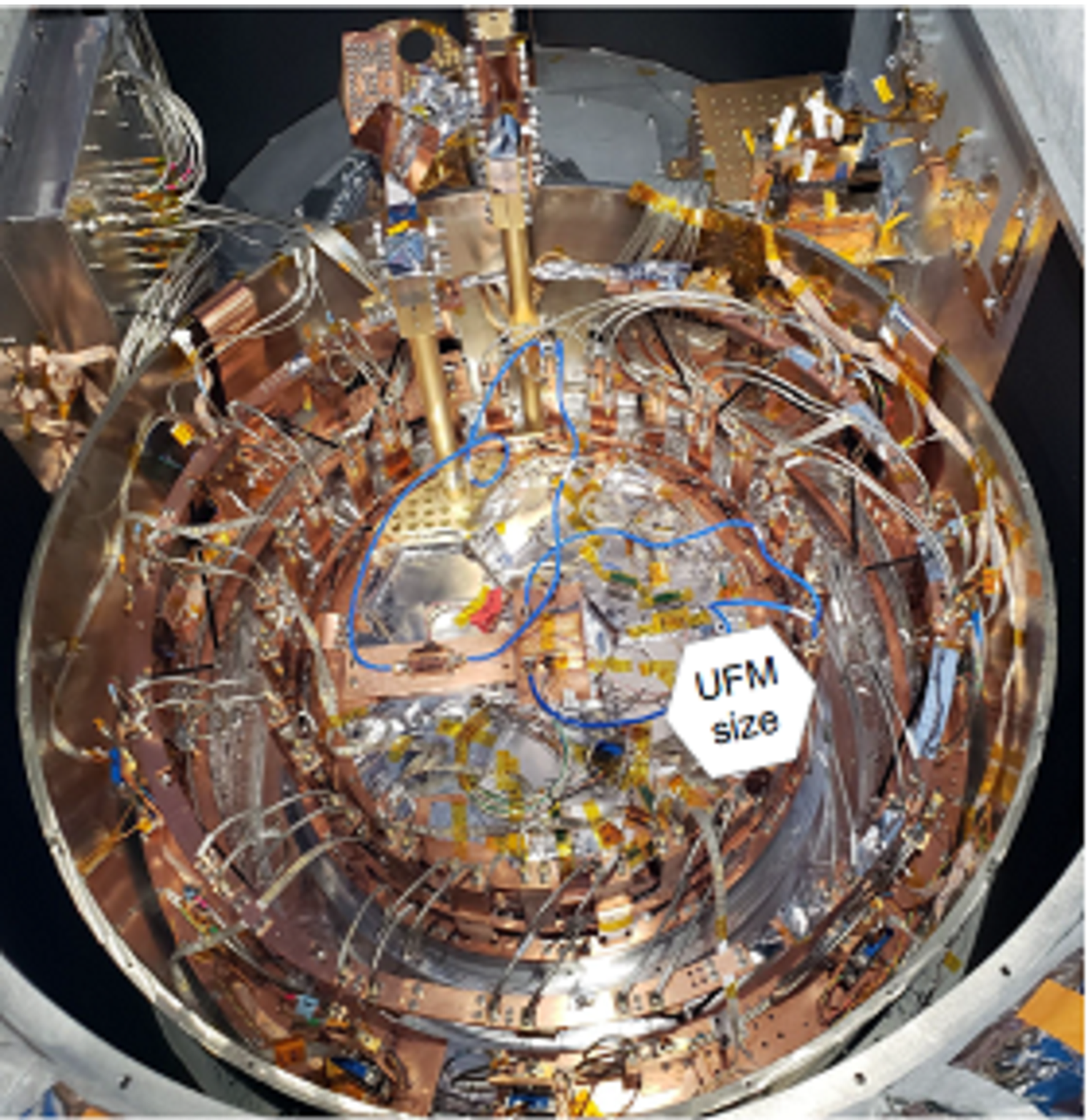}
    \end{tabular}
    \caption{$left$ picture is a rotational part of sparse wire-grid. $right$ picture is a cold readout assembly. The detector module (UFM) will be installed innermost part and coaxial wires and DC wires are connected to the outer side of cryostat.}
    \label{fig:wg_cra}
\end{figure}

Figure \ref{fig:wg_cra} $(left)$ shows the rotating part of the sparse wire-grid calibrator.
This is designed to calibrate the relative polarization angle.
It is mounts on top of the vacuum window.
The grid of thin tungsten wires produce uniform polarized signal and
polarized signal, and this signal passes through all of the optical components (the HWP, lenses, LPEs and etc) before it reaches detectors.
Typically 39 wires are aligned in one direction to within $0.1{}^\circ$ to calibrate the relative angle of all sensors with the optical components with $0.1{}^\circ$ level.
The wire ring is mounted on a circular bearing and can rotate in stepwise.
We will develop automatic loading system allowing us to calibrate the system frequently.

\subsection{Detector and Focal Plane} 
SO will use a common detector module, referred to as a universal focal-plane module (UFM) for the LAT and the SATs.
The detector module consists of 6-inch diameter TES wafers and multiplexing chips.
The operational temperature of our detector modules are designed and fabricated to be operated at a 100~mK stage temperature..
The TESes use Aluminum-Manganese (AlMn) alloy and tune the critical temperature to 160~mK.
Each pixel is dual polarization sensitive and di-chroic.
The microwave circuits including band-pass filters are integrated on a TES wafer,
which is fabricated using 6-inch photolithography process.
The feed-horn coupled TES array will cover MF and UHF\cite{Walker2020}, and lenslet coupled TES array will cover LF\cite{Suzuki2016}.
The target multiplexing factor of detector module is $\sim$ 1,000.
We adopt Microwave SQUID Multiplexer ($\mu$MUX \cite{Irwin2004,Mates2008}) that reads the TES using RF SQUID amplifier connected to a unique superconducting quarter wavelength resonators between 4-6~GHz\cite{umux,Erin2020}.
We inject a ramp wave through the flux ramp line that inductively couples to the RF SQUID loop and modulate signal.
This flux ramp modulation corrects the non-linearity of SQUID and helps mitigate 1/f noise in the resonator.
The SLAC Superconducting Microresonator RF electronics (SMuRF\cite{smurf}) will be used as a warm readout system for $\mu$MUX.
Each detector (TES with $\mu$MUX) with SMuRF readout is designed to be photon noise limited under out expected operating conditions in Chile.
Seven detector modules will be installed in one SAT focal plane, and more than 30,000 detectors will be operated in three SATs total.
Figure \ref{fig:wg_cra} $(right)$ shows a cold readout assembly including low noise HEMT amplifiers and coaxial wires from 4~K to 1~K can be assembled separately\cite{Rao2020}.
Thermal loading is enough smaller than the cooling power.
Thermal noise is designed to be smaller than a photon noise limited. 
The linearity of the amplifiers are well controlled for the high multiplexing factor of 1,000.

\section{Conclusion} 
The SAT telescopes are optimized to measure a B-mode signal produced by the primordial gravitational waves.
Our science goal is to measure this signal with $\sigma(r)=0.002$.
The SAT components are all designed and the SAT platform, the cryostat and the optics tube are already being built.
The development of HWP, calibrator, detector modules and readout are ongoing and mechanical parts for these instruments are fabricated.
We have begun integration in 2020 and aim to deploy the first SAT in Chile in 2021.

\acknowledgments     
This work was supported in part by a grant from the Simons Foundation (Award \#457687, B.K.)
This work was supported in part by the World Premier International Research Center Initiative (WPI Initiative), MEXT, Japan.
In Japan, this work was supported by JSPS KAKENHI grant Nos. 17H06134, 16K21744 and 19H00674, and the JSPS Core-to-Core Program JPJSCCA20200003.
Work at LBNL is supported in part by the U.S. Department of Energy, Office of Science,Office of High Energy Physics, under contract No. DE-AC02-05CH11231.
ZX is supported by the Gordon and Betty Moore Foundation

\bibliography{report}   

\begin{thebibliography}{10}

\bibitem{soforecast2019}
Ade, P., Aguirre, J., Ahmed, Z., Aiola, S., Ali, A., Alonso, D., Alvarez,
  M.~A., Arnold, K., Ashton, P., Austermann, J., and et~al., ``The simons
  observatory: science goals and forecasts,'' {\em Journal of Cosmology and
  Astroparticle Physics}~{\bf 2019},  056–056 (Feb 2019).

\bibitem{bmode1}
Kamionkowski, M., Kosowsky, A., and Stebbins, A., ``A probe of primordial
  gravity waves and vorticity,'' {\em Phys. Rev. Lett.}~{\bf 78},  2058--2061
  (Mar 1997).

\bibitem{bmode2}
Alonso, D., Dunkley, J., Thorne, B., and N\ae{}ss, S., ``Simulated forecasts
  for primordial $b$-mode searches in ground-based experiments,'' {\em Phys.
  Rev. D}~{\bf 95},  043504 (Feb 2017).

\bibitem{sodecadal2019}
{The Simons Observatory Collaboration} et~al., ``The simons observatory:
  Astro2020 decadal project whitepaper,'' (2019).

\bibitem{datta2013ApOpt..52.8747D}
{Datta}, R., {Munson}, C.~D., {Niemack}, M.~D., {McMahon}, J.~J., {Britton},
  J., {Wollack}, E.~J., {Beall}, J., {Devlin}, M.~J., {Fowler}, J., {Gallardo},
  P., {Hubmayr}, J., {Irwin}, K., {Newburgh}, L., {Nibarger}, J.~P., {Page},
  L., {Quijada}, M.~A., {Schmitt}, B.~L., {Staggs}, S.~T., {Thornton}, R., and
  {Zhang}, L., ``Large-aperture wide-bandwidth antireflection-coated silicon
  lenses for millimeter wavelengths,'' {\em Appl. Opt.}~{\bf 52},  8747 (Dec.
  2013).

\bibitem{nick2018}
Galitzki, N. et~al., ``{The Simons Observatory: instrument overview},'' in
  [{\em Millimeter, Submillimeter, and Far-Infrared Detectors and
  Instrumentation for Astronomy IX}{\nolinebreak\hspace{0.1em}]},  Zmuidzinas,
  J. and Gao, J.-R., eds.,  {\bf 10708},  1 -- 13, International Society for
  Optics and Photonics, SPIE (2018).

\bibitem{Joey2020}
Golec, J.~E. et~al., ``Design and fabrication metamaterail anti-reflectiuon
  coatings for cmb observations,'' International Society for Optics and
  Photonics, SPIE (2020).

\bibitem{xu2020arXiv201002233X}
{Xu}, Z., {Chesmore}, G.~E., {Adachi}, S., {Ali}, A.~M., {Bazarko}, A.,
  {Coppi}, G., {Devlin}, M., {Devlin}, T., {Dicker}, S.~R., {Gallardo}, P.~A.,
  {Golec}, J.~E., {Gudmundsson}, J.~E., {Harrington}, K., {Hattori}, M.,
  {Kofman}, A., {Kiuchi}, K., {Kusaka}, A., {Limon}, M., {Matsuda}, F.,
  {McMahon}, J., {Nati}, F., {Niemack}, M.~D., {Suzuki}, A., {Teply}, G.~P.,
  {Thornton}, R.~J., {Wollack}, E.~J., {Zannoni}, M., and {Zhu}, N., ``{The
  Simons Observatory: Metamaterial Microwave Absorber (MMA) and its Cryogenic
  Applications},'' {\em arXiv e-prints} ,  arXiv:2010.02233 (Oct. 2020).

\bibitem{adachi2020}
Adachi, S., Hattori, M., Kanno, F., Kiuchi, K., Okada, T., and Tajima, O.,
  ``Production method of millimeter-wave absorber with 3d-printed mold,'' {\em
  Review of Scientific Instruments}~{\bf 91}(1),  016103 (2020).

\bibitem{charlie2020arXiv200903972H}
{Hill}, C.~A., {Kusaka}, A., {Ashton}, P., {Barton}, P., {Adkins}, T.,
  {Arnold}, K., {Bixler}, B., {Ganjam}, S., {Lee}, A.~T., {Matsuda}, F.,
  {Matsumura}, T., {Sakurai}, Y., {Tat}, R., and {Zhou}, Y., ``{A cryogenic
  continuously rotating half-wave plate for the POLARBEAR-2b cosmic microwave
  background receiver},'' {\em arXiv e-prints} ,  arXiv:2009.03972 (Sept.
  2020).

\bibitem{Walker2020}
Walker, S., Sierra, C.~E., Austermann, J.~E., Beall, J.~A., Becker, D.~T.,
  Dober, B.~J., Duff, S.~M., Hilton, G.~C., Hubmayr, J., Van~Lanen, J.~L.,
  McMahon, J.~J., Simon, S.~M., Ullom, J.~N., and Vissers, M.~R.,
  ``Demonstration of 220/280 ghz multichroic feedhorn-coupled tes
  polarimeter,'' {\em Journal of Low Temperature Physics}~{\bf 199},  891--897
  (May 2020).

\bibitem{Suzuki2016}
Suzuki, A. et~al., ``The polarbear-2 and the simons array experiments,'' {\em
  Journal of Low Temperature Physics}~{\bf 184},  805--810 (Aug 2016).

\bibitem{Irwin2004}
Irwin, K.~D. and Lehnert, K.~W., ``Microwave squid multiplexer,'' {\em Applied
  Physics Letters}~{\bf 85}(11),  2107--2109 (2004).

\bibitem{Mates2008}
Mates, J. A.~B., Hilton, G.~C., Irwin, K.~D., Vale, L.~R., and Lehnert, K.~W.,
  ``Demonstration of a multiplexer of dissipationless superconducting quantum
  interference devices,'' {\em Applied Physics Letters}~{\bf 92}(2),  023514
  (2008).

\bibitem{umux}
Dober, B., Ahmed, Z., Becker, D.~T., Bennett, D.~A., Connors, J.~A., Cukierman,
  A., D'Ewart, J.~M., Duff, S.~M., Dusatko, J.~E., Frisch, J.~C., Gard, J.~D.,
  Henderson, S.~W., Herbst, R., Hilton, G.~C., Hubmayr, J., Mates, J. A.~B.,
  Reintsema, C.~D., Ruckman, L., Ullom, J.~N., Vale, L.~R., Winkle, D. D.~V.,
  Vasquez, J., Young, E., and Yu, C., ``A microwave squid multiplexer optimized
  for bolometric applications,'' (2020).

\bibitem{Erin2020}
Hearly, E. et~al., ``Assembly development for the simons observatory focal
  plane readout module,'' International Society for Optics and Photonics, SPIE
  (2020).

\bibitem{smurf}
Henderson, S.~W., Ahmed, Z., Austermann, J., Becker, D., Bennett, D.~A., Brown,
  D., Chaudhuri, S., Cho, H.-M.~S., D'Ewart, J.~M., Dober, B., Duff, S.~M.,
  Dusatko, J.~E., Fatigoni, S., Frisch, J.~C., Gard, J.~D., Halpern, M.,
  Hilton, G.~C., Hubmayr, J., Irwin, K.~D., Karpel, E.~D., Kernasovskiy, S.~S.,
  Kuenstner, S.~E., Kuo, C.-L., Li, D., Mates, J. A.~B., Reintsema, C.~D.,
  Smith, S.~R., Ullom, J., Vale, L.~R., Winkle, D. D.~V., Vissers, M., and Yu,
  C., ``{Highly-multiplexed microwave SQUID readout using the SLAC
  Microresonator Radio Frequency (SMuRF) electronics for future CMB and
  sub-millimeter surveys},'' in [{\em Millimeter, Submillimeter, and
  Far-Infrared Detectors and Instrumentation for Astronomy
  IX}{\nolinebreak\hspace{0.1em}]},  Zmuidzinas, J. and Gao, J.-R., eds.,  {\bf
  10708},  170 -- 185, International Society for Optics and Photonics, SPIE
  (2018).

\bibitem{Rao2020}
Sathyanarayana~Rao, M., Silva-Feaver, M., Ali, A., Arnold, K., Ashton, P.,
  Dober, B.~J., Duell, C.~J., Duff, S.~M., Galitzki, N., Healy, E., Henderson,
  S., Ho, S.-P.~P., Hoh, J., Kofman, A.~M., Kusaka, A., Lee, A.~T., Mangu, A.,
  Mathewson, J., Mauskopf, P., McCarrick, H., Moore, J., Niemack, M.~D., Raum,
  C., Salatino, M., Sasse, T., Seibert, J., Simon, S.~M., Staggs, S., Stevens,
  J.~R., Teply, G., Thornton, R., Ullom, J., Vavagiakis, E.~M., Westbrook, B.,
  Xu, Z., and Zhu, N., ``Simons observatory microwave squid multiplexing
  readout: Cryogenic rf amplifier and coaxial chain design,'' {\em Journal of
  Low Temperature Physics}~{\bf 199},  807--816 (May 2020).

\end{thebibliography}
\bibliographystyle{spiebib}   

\end{document}